\documentclass[onecolumn,aps,prd,superscriptaddress,preprintnumbers,nofootinbib,10pt]{revtex4-2}
\usepackage{amsfonts}
\usepackage[tbtags]{amsmath}
\usepackage{amssymb}
\usepackage{color}
\usepackage{courier}
\usepackage{dsfont}
\usepackage{euscript}
\usepackage{epsfig}
\usepackage{epstopdf}
\usepackage{float}
\usepackage{graphics}
\usepackage{subcaption}
\usepackage{graphicx}
\usepackage{hyperref}
\usepackage[utf8]{inputenc}
\usepackage{latexsym}
\usepackage{MnSymbol}
\usepackage{pifont}
\usepackage{physics}
\usepackage{slashed}
\usepackage{tikz-feynman}
\usepackage[normalem]{ulem}
\usepackage{url}
\usepackage{verbatim}
\usepackage{widetable}
\usepackage{setspace}

\allowdisplaybreaks

\hypersetup{
	colorlinks=true,
	citecolor=blue,
	citebordercolor=red,
	linktoc=all,
	linkcolor=blue,
	urlcolor=blue
}

\begin{document}	
	
	\title{{\bf \Large  Cat states and violation of the Bell-CHSH inequality \\in relativistic Quantum Field Theory}}
	
	\vspace{1cm}

	\author{M. S.  Guimaraes}\email{msguimaraes@uerj.br} \affiliation{UERJ -- Universidade do Estado do Rio de Janeiro,	Instituto de Física -- Departamento de Física Teórica -- Rua São Francisco Xavier 524, 20550-013, Maracanã, Rio de Janeiro, Brazil}
	
	\author{I. Roditi} \email{roditi@cbpf.br} \affiliation{CBPF $-$ Centro Brasileiro de Pesquisas Físicas, Rua Dr. Xavier Sigaud 150, 22290-180, Rio de Janeiro, Brazil } 
	
	\author{S. P. Sorella} \email{silvio.sorella@fis.uerj.br} \affiliation{UERJ -- Universidade do Estado do Rio de Janeiro,	Instituto de Física -- Departamento de Física Teórica -- Rua São Francisco Xavier 524, 20550-013, Maracanã, Rio de Janeiro, Brazil}

	\begin{abstract}

A cat state localized in the right Rindler wedge is employed to study the violation of the Bell-CHSH inequality in a relativistic scalar free Quantum Field Theory. By means of the bounded Hermitian operator  $sign(\varphi(f))$, where $\varphi(f)$ stands for the smeared scalar field, it turns out that the Bell-CHSH correlator can be evaluated in closed analytic form in terms of the imaginary error function. Being the superposition of two coherent states, cat states allow for the existence of interference terms which give rise to a violation of the Bell-CHSH inequality. As such, the present setup can be considered as an explicit realization of the results obtained by Summers-Werner.

		\end{abstract}
		
		\maketitle

\section{Introduction}\label{sect1}

In a series of seminal papers \cite{Summers:1987fn,Summers:1987squ,Summers:1987ze}, Summers-Werner have provided a rigorous framework for the Bell-CHSH inequality \cite{Bell:1964kc,Clauser:1969ny}  in relativistic Quantum Field Theory, for both Bose and Fermi fields. They have been able to establish a set of theorems stating that the Bell-CHSH inequality achieves maximal violation already at the level of free fields, for judicious observables located in complementary wedge regions as well as in double tangent cones. These results were established first for the vacuum state \cite{Summers:1987fn,Summers:1987squ} and later on extended to a broader class of  states \cite{Summers:1987ze}\footnote{See \cite{Guimaraes:2024byw,Guimaraes:2024mmp} for recent reviews of the Bell-CHSH inequality in Quantum Mechanics and Quantum Field Theory.}. \\\\The aforementioned theorems rely on fundamental aspects of Quantum Field Theory such as: the Haag-Kastler algebraic formulation  \cite{Haag:1992hx}, the Reeh-Schlieder \cite{Reeh:1961ujh} and Bisognano-Wichmann \cite{Bisognano:1975ih} theorems, the theory of von Neumann algebras \cite{Haag:1992hx,Witten:2018zxz} and the modular theory of Tomita-Takesaki \cite{Takesaki:1970aki,Bratteli:1979tw,Summers:2003tf,Guido:2008jk,Witten:2018zxz}. \\\\Nevertheless, it seems fair to state that addressing the Bell-CHSH inequality in Quantum Field Theory is a rather challenging topic. The effective implementation of the Summers-Werner results into a computational framework still deserves much work. For instance, in the case of a scalar field, the explicit construction of the bounded operators whose correlation functions saturate Tsirelson's bound \cite{Cirelson:1980ry} has not yet been worked out. \\\\In recent years we have devoted efforts towards a realization of the Summers-Werner theorems, searching for an example which could enable for an evaluation of the Bell-CHSH correlator in closed analytic form. Before going any further, a short overview of what we have been able to obtain is in order. In \cite{DeFabritiis:2024jfy,Guimaraes:2024xtj} a numerical setup for the Bell-CHSH inequality has been outlined for both wedge regions and tangent double cones. In \cite{DeFabritiis:2023tkh}, the modular theory of Tomita-Takesaki has been used to study the properties of the correlation functions of unitary Weyl operators. Finally, in \cite{Guimaraes:2024alk,Guimaraes:2025vfu}, we have looked at the difficult issue of constructing a set of bounded Hermitian operators useful to investigate the Bell-CHSH inequality. \\\\In this new contribution, we present a concrete  example of a Bell-CHSH correlation function which can be evaluated in closed form by making use of the Summers-Werner theorems, while exhibiting a clean violation of the classical bound; {\it i.e.} $2$. \\\\In order to illustrate the salient points of the work, let us remind the expression of the Bell-CHSH correlator of a scalar Quantum Field Theory, namely: 
\begin{equation} 
\langle {\cal C} \rangle_{\psi} = \langle \psi|  (A(f) +A(f'))B(g) + (A(f)-A(f'))B(g')  |\psi \rangle \;, \label{ineq}
\end{equation}
where $(A(f),A(f'),B(g),B(g'))$ is a shorthand notation for 
\begin{equation} 
A(f) = A(\varphi(f))\;, \qquad A(f') = A(\varphi(f'))\;, \qquad B(g) = B(\varphi(g))\;, \qquad B(g') = B(\varphi(g'))\;, \label{short}
\end{equation}
with $\varphi(t,{\vec x})$ denoting the scalar field and $\varphi(f)$ its smearing \cite{Haag:1992hx}, see Appendix \eqref{App}.
\begin{equation} 
\varphi(f) = \int d^4x\; \varphi(x) f(x) \;, \label{smf}
\end{equation}
where $f(t,{\vec x})$ is a smooth, compactly supported, test function. \\\\The operators $(A(f),A(f'))$ and $(B(g),B(g'))$ are Hermitian bounded operators
\begin{equation} 
||A(f)|| \le 1\;, \qquad ||A(f')|| \le 1\;, \qquad ||B(g)|| \le 1\;, \qquad ||B(g')|| \le 1\;,  \label{bound}
\end{equation}
fulfilling the following conditions 
\begin{eqnarray} 
\left[A(f),B(g) \right] & =& 0 \;, \qquad [A(f),B(g')]=0 \;,  \qquad [A(f),A(f')] \neq 0 \;, \nonumber \\
\left[A(f'),B(g)\right] & =& 0 \;, \qquad [A(f'),B(g')]=0 \;,  \qquad [B(g),B(g')] \neq 0 \;, \label{Bella}
\end{eqnarray} 
meaning that $(A(f),A(f'))$ are space-like separated  with respect to $(B(g),B(g'))$. According to \cite{Summers:1987fn,Summers:1987squ,Summers:1987ze}, the Bell-CHSH inequality is said to be violated whenever 
\begin{equation} 
2 < \big|  \langle {\cal C} \rangle_{\psi} \big| \le 2 \sqrt{2} \;. \label{viol}
\end{equation} 
As one can see from equation \eqref{ineq}, three key ingredients are required for the Bell-CHSH correlator in Quantum Field Theory: the state $|\psi\rangle$, the test functions $(f,f')$, $(g,g')$ and the operators $(A(f),A(f'),B(g),B(g'))$ . These issues will be addressed in the following way: 
\begin{itemize} 
\item the state $|\psi\rangle$ will be  chosen to be a cat state localized in the right wedge ${\cal O}_R= \{(t,x,y,z)\;, x\ge |t| \}$\footnote{Similarly, the left wedge is defined by ${\cal O}_L= \{(t,x,y,z)\;, -x\ge |t| \}$.}:
\begin{equation} 
|\psi\rangle = {\cal N} \left( e^{i \varphi(h)} + e^{i \sigma} e^{-i \varphi(h)} \right) |0\rangle \;, \label{cat}
\end{equation}
where $|0\rangle$ is the vacuum state, ${\cal N}$ a normalization factor, $\sigma$ an arbitrary phase and $h$ a suitable test function supported in ${\cal O}_R$. The state $|\psi\rangle$ is the superposition of two coherent states obtained by applying the Weyl operator $W_h = e^{i \varphi(h)}$ to the vacuum state $|0\rangle$. Being a superposition, the cat state $|\psi\rangle$ gives rise to interference terms in the correlation function \eqref{ineq} which, as we shall see, will play a crucial role for the violation of the Bell-CHSH inequality. 

\item the two sets of space-like test functions $(f,f')$ and $(g,g')$ will be specified by strictly following the original construction outlined in \cite{Summers:1987squ}\footnote{See Proposition 3.1 of \cite{Summers:1987squ}.}, relying on the Tomita-Takesaki modular theory as well as on the knowledge of the spectrum of the modular operator in the wedge regions $({\cal O}_R,{\cal O}_L)$  \cite{Bisognano:1975ih}. As such, $(f,f')$ and $(g,g')$ encapsulate fundamental aspects of both modular theory and of the type III nature of the von Neumann algebras of Quantum Field Theory  \cite{Summers:2003tf,Guido:2008jk,Witten:2018zxz}. 

\item concerning now the choice of operators, we shall rely on the construction outlined in \cite{Guimaraes:2024alk,Guimaraes:2025vfu} and consider operators obtained as continuous superpositions of Weyl operators. In particular, we shall focus on the operator $sign(\varphi(f))$, defined through the functional extension of the Dirichlet integral representation of the $ sign$ function:

\begin{equation}
A(f)=sign(\varphi(f)) = \frac{2}{\pi} \int_0^\infty dk\; \frac{1}{k}\; \sin(k \varphi(f)) = \frac{1}{i\pi} \int_0^\infty dk\; \frac{1}{k}\;\left( e^{ik \varphi(f)} - e^{-ik \varphi(f)} \right) \;. \label{sgn}
\end{equation}
Analogously, we set 
\begin{equation} 
A(f')=sign(\varphi(f'))\;, \qquad B(g)=sign(\varphi(g))\;, \qquad B(g')=sign(\varphi(g')) \;. \label{opsgn}
\end{equation} 
The operators \eqref{sgn},\eqref{opsgn} fulfill the conditions \eqref{bound},\eqref{Bella}, being eligible for the Bell-CHSH inequality. The usefulness of employing the operators \eqref{sgn},\eqref{opsgn} relies on the fact that the Bell-CHSH correlator, eq.\eqref{ineq}, can be evaluated in analytic closed form, a rare feature in Quantum Field Theory. 

\end{itemize} 
The paper is organized as follows. In Sect.\eqref{sec2} we evaluate the correlation function of the operators \eqref{sgn},\eqref{opsgn} needed for the Bell-CHSH inequality. The computation will be done for a generic choice of the test functions $(f,f')$, $(g,g')$ and $h$, eqs.\eqref{cat},\eqref{sgn},\eqref{opsgn}. In Sect.\eqref{sec3} we specialize to the Summers-Werner test functions and discuss the ensuing violation of the Bell-CHSH inequality. Sect.\eqref{sec4} summarizes our conclusion. For the benefit of the reader and in order to present the material in a self-contained way, in Appendix \eqref{App} we remind a few basic notions of the canonical quantization of the scalar field and of the Tomita-Takesaki modular theory.

\section{Evaluation of the correlation functions of the sign operators}\label{sec2}

Let us proceed by evaluating the correlation function of the $sign$ operators, eqs.\eqref{sgn},\eqref{opsgn}: 
\begin{equation} 
\langle \psi |\; A(f) B(g) \;|\psi\rangle = \langle \psi |\; sign(\varphi(f)) \;sign(\varphi(g)) \;|\psi \rangle \;, \label{crs}
\end{equation}
where $\varphi(f)$ is the smeared field,  see Appendix \eqref{App}, and $|\psi\rangle$ is the cat state of eq.\eqref{cat}. The normalization factor ${\cal N}$ is easily computed from $\langle \psi|\psi\rangle=1$, being given by 
\begin{equation} 
{\cal N}^2 =  \frac{1}{2\left( 1+ \cos(\sigma) e^{-2 ||h||^2}\right)} \;, \label{nn}
\end{equation} 
where $||h||^2 = \langle h|h \rangle$ is the norm of the test function $h$, eqs.\eqref{Inn}-\eqref{fin}. Therefore 
\begin{eqnarray} 
\langle \psi |\; A(f) B(g) \;|\psi\rangle & =& -\frac{{\cal N}^2}{\pi^2} \int_0^\infty \frac{dk dq}{k q}\; \times \nonumber \\
&\; & \langle 0| \; \left( e^{-i \varphi(h)} + e^{-i \sigma} e^{i \varphi(h)} \right) \left( e^{i k\varphi(f)} - e^{-i k \varphi(f)} \right)  \left( e^{i \varphi(h)} + e^{i \sigma} e^{-i \varphi(h)} \right) \left( e^{i q\varphi(g)} - e^{-i q \varphi(g)} \right)\;|0\rangle \;, \nonumber \\ \label{corr1}
\end{eqnarray}
where use has been made of the fact that $(f,f',h)$ are supported in the right wedge ${\cal O}_R$ while $(g,g')$ in the left one  ${\cal O}_L$: 
\begin{eqnarray} 
\left[\varphi(f), \varphi(g)\right] & = & \left[\varphi(f'), \varphi(g)\right]=\left[\varphi(h), \varphi(g)\right] =0 \;, \nonumber \\
\left[\varphi(f), \varphi(g')\right] & = & \left[\varphi(f'), \varphi(g')\right]=\left[\varphi(h), \varphi(g')\right] =0 \;. \label{cmmt}
\end{eqnarray}
Using then the properties of the Weyl operators, eq.\eqref{algebra}, it follows that, after some algebra, 
\begin{eqnarray} 
\langle \psi |\; A(f) B(g) \;|\psi\rangle & =& -\frac{{\cal N}^2}{\pi^2} \int_0^\infty \frac{dk dq}{k q}\;  \langle0| \; \left( e^{i q \varphi(g)} - e^{-iq  \varphi(g)} \right) \times \nonumber \\
&\; & \left[ e^{i k \varphi(f)} e^{-i k \Delta_{PJ}(f,h)} + e^{i \sigma} e^{i(k \varphi(f) - 2 \varphi(h))} - e^{-i k \varphi(f)} e^{i k \Delta_{PJ}(f,h)} -e^{i \sigma} e^{-i(k \varphi(f) + 2 \varphi(h))} \right.  \nonumber \\
&+& \left. e^{i k \varphi(f)} e^{i k \Delta_{PJ}(f,h)} + e^{-i \sigma} e^{i(k \varphi(f) + 2 \varphi(h))} - e^{-i k \varphi(f)} e^{-i k \Delta_{PJ}(f,h)} -e^{-i \sigma} e^{-i(k \varphi(f) - 2 \varphi(h))} \right.] \; |0\rangle \;, \nonumber \\
\label{crrt}
\end{eqnarray}
where $\Delta_{PJ}(f,h)$ stands for the smeared Pauli-Jordan distribution, eqs.\eqref{mint},\eqref{PJH}. Therefore, from eqs.\eqref{vA},\eqref{vAhh}, it follows 
\begin{eqnarray} 
\langle \psi |\; A(f) B(g) \;|\psi\rangle & =& -\frac{{\cal N}^2}{\pi^2} \int_0^\infty \frac{dk dq}{k q}\;  \left( - 8 \cos(k \Delta_{PJ}(f,h)) e^{-\frac{1}{2}(k^2 ||f||^2 + q^2 ||g||^2)}\sinh(kq H(f,g)) \right. \nonumber \\
&+ & \left. 4 \cos(\sigma) e^{-\frac{1}{2}||kf + qg||^2 } e^{-2 ||h||^2} \cosh(H(kf+qg,h)) - 
4 \cos(\sigma) e^{-\frac{1}{2}||kf - qg||^2 } e^{-2 ||h||^2} \cosh(H(kf-qg,h)) \right) \nonumber \\
\label{crrt1}
\end{eqnarray}
where $H(f,g)$ is the smeared Hadamard distribution, eqs.\eqref{mint},\eqref{PJH}.\\\\Finally, for the Bell-CHSH correlation function, one has 
\begin{eqnarray} 
\langle \psi |\; {\cal C} \;|\psi\rangle & =& \frac{2}{\pi^2} \frac{1}{1+ \cos(\sigma) e^{-2 ||h||^2}} \int_0^\infty \frac{dk dq}{k q}\;  \left( 2 \cos(k \Delta_{PJ}(f,h)) e^{-\frac{1}{2}(k^2 ||f||^2 + q^2 ||g||^2)}\sinh(kq H(f,g)) \right. \nonumber \\
&+ &2\cos(k \Delta_{PJ}(f',h)) e^{-\frac{1}{2}(k^2 ||f'||^2 + q^2 ||g||^2)}\sinh(kq H(f',g)) + 2  \cos(k \Delta_{PJ}(f,h)) e^{-\frac{1}{2}(k^2 ||f||^2 + q^2 ||g'||^2)}\sinh(kq H(f,g')) \nonumber \\
&-& 2  \cos(k \Delta_{PJ}(f',h)) e^{-\frac{1}{2}(k^2 ||f'||^2 + q^2 ||g'||^2)}\sinh(kq H(f',g)) \nonumber \\
&- & \cos(\sigma) e^{-\frac{1}{2}||kf + qg||^2 } e^{-2 ||h||^2} \cosh(H(kf+qg,h)) +
 \cos(\sigma) e^{-\frac{1}{2}||kf - qg||^2 } e^{-2 ||h||^2} \cosh(H(kf-qg,h)) \nonumber \\
&- &  \cos(\sigma) e^{-\frac{1}{2}||kf' + qg||^2 } e^{-2 ||h||^2} \cosh(H(kf'+qg,h)) + 
 \cos(\sigma) e^{-\frac{1}{2}||kf' - qg||^2 } e^{-2 ||h||^2} \cosh(H(kf'-qg,h))  \nonumber \\
&- & \cos(\sigma) e^{-\frac{1}{2}||kf + qg'||^2 } e^{-2 ||h||^2} \cosh(H(kf+qg',h)) +
 \cos(\sigma) e^{-\frac{1}{2}||kf - qg'||^2 } e^{-2 ||h||^2} \cosh(H(kf-qg',h))  \nonumber \\
 &+& \left. \cos(\sigma) e^{-\frac{1}{2}||kf' + qg'||^2 } e^{-2 ||h||^2} \cosh(H(kf'+qg',h)) -
\cos(\sigma) e^{-\frac{1}{2}||kf' - qg'||^2 } e^{-2 ||h||^2} \cosh(H(kf'-qg',h)) \right) \nonumber \\
\label{CC}
\end{eqnarray}

\section{The Summers-Werner inner products and the violation of the Bell-CHSH inequality}\label{sec3}	

\begin{figure}[t!]
\begin{minipage}[b]{0.4\linewidth}
\includegraphics[width=\textwidth]{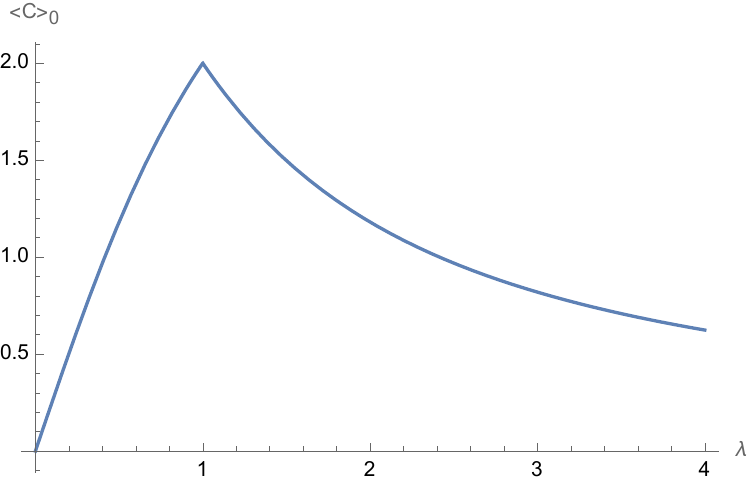}
\end{minipage} \hfill
\caption{Behavior of the correlator$ \langle {\cal C}\rangle_0$  as a function of the spectral parameter $\lambda$. Although $\lambda$ belongs to the spectral interval $[0,1]$, the plot has been extended over a larger interval to show that $\langle  {\cal C} \rangle_0$ is bounded by 2.  }
\label{plot}
\end{figure}

\begin{figure}[t!]
\begin{minipage}[b]{0.4\linewidth}
\includegraphics[width=\textwidth]{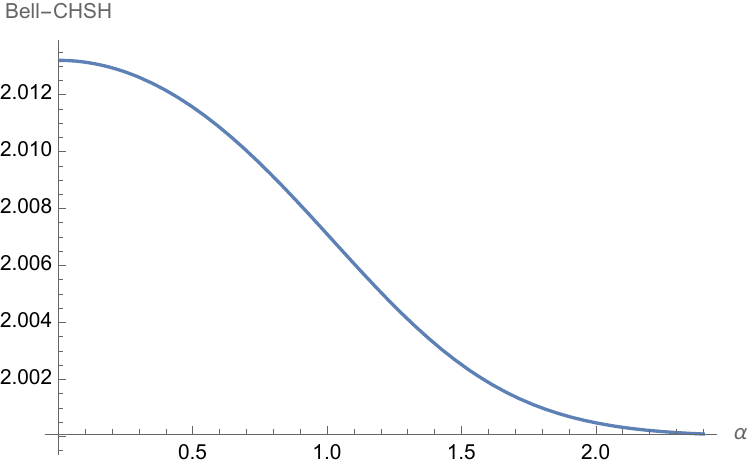}
\end{minipage} \hfill
 \caption{Behavior of the  correlator $ \langle {\cal C}\rangle_{\psi}$  as  function of the parameter $\alpha$, showing the violation of the Bell-CHSH inequality. The  value 2 is asymptotically recovered for very large values of $\alpha$. }
\label{catf}
\end{figure}

Having evaluated the correlation functions of the $sign$ operators in the cat state $|\psi\rangle$, eqs.\eqref{sgn},\eqref{opsgn}, we can now face the violation of the Bell-CHSH inequality. The first task is that of specifying the test functions $(f,f')$, $(g,g')$ and $h$. Here, we shall rely on the construction of \cite{Summers:1987squ}. From Proposition 3.1 of \cite{Summers:1987squ}, it follows that the test functions $(f,f',g,g')$ can be chosen in such a way that: 
\begin{eqnarray} 
\vert\vert f \vert\vert^2  &=& \vert\vert f'  \vert\vert^2 = \vert\vert g  \vert\vert^2 =\vert\vert g'  \vert\vert^2 = \frac{1+\lambda^2}{1-\lambda^2}
 \;, \nonumber \\
\langle f \vert g  \rangle &=& - \langle f' \vert g'  \rangle=   \frac{2 \lambda}{1-\lambda^2} \;,  \nonumber \\
\langle f \vert g'  \rangle &=&\langle f' \vert g  \rangle = 0 \;, \nonumber \\
\langle f \vert f'  \rangle &=&\langle g \vert g'  \rangle = i \;. 
  \label{ssffgg}
\end{eqnarray} 
where the parameter $\lambda$ belongs to the spectral interval $[0,1]$ of the Tomita-Takesaki modular operator \cite{Summers:1987squ}, see Appendix \eqref{App} for more details. Reminding that, as $(f,f')$, the test function $h$ is supported in the right wedge region ${\cal O}_R$, we shall set 
\begin{equation} 
h = \frac{\alpha}{\sqrt{2}} \frac{\sqrt{1-\lambda^2}}{\sqrt{1+\lambda^2}} (f+f') \;, \label{htest}
\end{equation}
with $\alpha$ an arbitrary parameter, to be chosen at the best convenience. Expression \eqref{htest} implies that 
\begin{equation} 
||h||^2 = \alpha^2 \;. \label{hnnmm}
\end{equation}
Plugging eqs.\eqref{ssffgg},\eqref{htest} into the correlation function \eqref{CC}, one gets 
\begin{equation} 
\langle \psi |\; {\cal C} \;|\psi\rangle  = \langle   {\cal C}  \rangle_0 + {\cal R} \;, \label{CC1}
\end{equation}
where $\langle  {\cal C} \rangle_0 $ denotes the expectation value of the operator ${\cal C}$ in the vacuum state, namely 
\begin{equation} 
\langle  {\cal C} \rangle_0 = \langle 0 |\; (A(f) + A(f'))B(g)+ (A(f)-A(f'))B(g') \;| 0\rangle = \frac{8}{\pi^2} \int_0^\infty \frac{dk dq }{kq} \; e^{-\frac{k^2}{4}} e^{-\frac{q^2}{4}}\; \sinh\left(\frac{kq \lambda}{1+\lambda^2} \right) \;. \label{cco}
\end{equation}		
For the remaining term ${\cal R}$ one has 
\begin{eqnarray} 
{\cal R} &= & \frac{1}{1+\cos(\sigma) e^{-2 \alpha^2}} \frac{8}{\pi^2} \int_0^\infty \frac{dk dq}{kq}\; e^{-\frac{1}{4}(k^2+q^2)}\left( 
\cos\left(  k \alpha \frac{1-\lambda^2}{1+\lambda^2} \right) \sinh\left( \frac{kq \lambda}{1+\lambda^2}\right)  -   (1+\cos(\sigma) e^{-2 \alpha^2}) \sinh\left(\frac{kq \lambda}{1+\lambda^2}\right)  \right. \nonumber \\
&+& \cos(\sigma) e^{-2 \alpha^2} \left(  \cosh(k \alpha) \cosh\left( \frac{2 q \alpha \lambda}{1+\lambda^2} \right)  \sinh\left(\frac{kq \lambda}{1+\lambda^2}\right) - \sinh(k \alpha) \sinh\left( \frac{2 q \alpha \lambda}{1+\lambda^2} \right)  \cosh\left(\frac{kq \lambda}{1+\lambda^2}\right)
 \right)  \;. \label{caca}
\end{eqnarray}
As it is easily figured out, the term ${\cal R}$ arises from the interference terms showing up from the superposition of the two coherent 
states, $e^{i \phi(h)} |0\rangle$ and $e^{-i \phi(h)} |0\rangle$, which give rise to the cat state $|\psi\rangle$.\\\\Moreover, recalling that the imaginary error function ${\it erfi}(w)$ can be represented as 
 \begin{equation} 
 {\it erfi}(w) = \frac{2}{\pi} \int_0^\infty \frac{dt}{t} e^{-\frac{t^2}{4}} \; \sinh(w t) \;, \label{erfi} 
 \end{equation}   
 it turns out that $\langle  {\cal C}\rangle_0$ can be cast in closed form as 
 \begin{equation} 
  \langle  {\cal C} \rangle_0 = \frac{4}{\pi}  \arcsin\left( \frac{2 \lambda}{1+\lambda^2}\right) \;. \label{cls}
 \end{equation} 
 Unfortunately, this expression is bounded by the classical value 2, as it can be seen from Fig.~\eqref{plot}. 
Notice that, according to \cite{Summers:1987squ}, the maximum value of  $\langle  {\cal C} \rangle_0$ occurs for $\lambda=1$. \\\\It remains now to work out the contribution of the term $\cal R$ in eq.\eqref{caca}. Setting $\lambda=1$, one obtains 
\begin{eqnarray}
{\cal R}_{\lambda=1}  &= & \frac{\cos(\sigma) e^{-2 \alpha^2}}{1+\cos(\sigma) e^{-2 \alpha^2}} \frac{8}{\pi^2} \int_0^\infty \frac{dk dq}{kq}\; e^{-\frac{1}{4}(k^2+q^2)}\left( \sinh\left( \frac{kq}{2} \right) \left(-1 + \cosh(k \alpha) \cosh(q \alpha)      \right) \right.
\nonumber \\
&- &
\left. \sinh\left( \frac{kq}{2} \right) \sinh(k \alpha) \sinh(q \alpha) \right)
\end{eqnarray}
Again, making use of the imaginary error function, the final expression for the Bell-CHSH correlation function simplifies to 
\begin{equation} 
\langle \psi |\; {\cal C} \;|\psi\rangle  = 2 + \frac{2}{\pi}\frac{\cos(\sigma) e^{-2 \alpha^2}}{1+\cos(\sigma) e^{-2 \alpha^2}}  {\cal J}(\alpha)  \;, \label{BCHSH-final}
\end{equation}
with 
\begin{equation}
{\cal J}(\alpha)  = \int_0^\infty \frac{dk}{k} e^{-\frac{k^2}{4}} \left( -2 erfi\left( \frac{k}{2} \right)  + e^{-k \alpha} erfi\left( \frac{k}{2}+ \alpha\right) +\cosh(\alpha k) erfi\left( \frac{k}{2}- \alpha\right) - \sinh(\alpha k) erfi\left( -\frac{k}{2}+ \alpha\right)
\right) \label{JJ}
\end{equation}    
Expressions \eqref{BCHSH-final},\eqref{JJ} represent our final result. They provide a very concise expression for the Bell-CHSH correlator depending on the parameters $(\alpha,\sigma)$ of the cat state $|\psi\rangle$. The integral ${\cal J}(\alpha)$ in eq.\eqref{JJ} can be  investigated through numerical tools, being a negative function of the parameter $\alpha$. We then choose $\sigma=\pi$, so that the second term in the right hand side of \eqref{BCHSH-final} becomes positive, yielding a clean proof of the violation of the Bell-CHSH inequality: 
\begin{equation} 
|\langle \psi |\; {\cal C} \;|\psi\rangle | > 2 \;. \label{violt-f}
\end{equation}
The size of the violation is of about $2.012$, as one can see from Fig.\eqref{catf}, where the behavior of $ |\langle \psi |\; {\cal C} \;|\psi\rangle |$ as a function of $\alpha$ is reported. 

\section{Conclusion} \label{sec4}

Elaborating on the seminal work of Summers-Werner \cite{Summers:1987fn,Summers:1987squ,Summers:1987ze}, we have been able to construct  a setup for the violation of the Bell-CHSH inequality in a relativistic scalar Quantum Field Theory. \\\\Three key ingredients have been employed, as enlisted below:
\begin{itemize} 
\item the state $|\psi\rangle$ has been chosen to be a cat state, {\it i.e.} a superposition of two coherent states, eq.\eqref{cat}. As such, cat states enable for the existence of interference terms which play a primary role for the violation of the Bell-CHSH inequality. 

\item the test functions $(f,f')$, $(g,g')$ and $h$ have been selected by strictly relying on the theorems established by Summers-Werner. These test functions follow from the knowledge of the properties of the Tomita-Takesaki modular operator in wedge regions \cite{Summers:1987fn,Summers:1987squ,Summers:1987ze}. 

\item finally, as bounded Hermitian operators, the $sign$ operators of eqs.\eqref{sgn},\eqref{opsgn} have been employed. These operators fulfill the conditions  \eqref{bound},\eqref{Bella}, being eligible for the Bell-CHSH inequality. 

\end{itemize}
These three ingredients have made it possible to cast the Bell-CHSH correlation function in closed concise form, as shown by eqs.\eqref{BCHSH-final},\eqref{JJ}, exhibiting a clean proof of the violation of the Bell-CHSH inequality in relativistic Quantum Field Theory. As such, the present example offers an explicit realization of the Summers-Werner results.

\section*{Acknowledgments}
The authors would like to thank the Brazilian agencies CNPq, CAPES and FAPERJ for financial support.  S. P.~Sorella, I.~Roditi, and M. S.~Guimaraes are CNPq researchers under contracts 302991/2024-7, 311876/2021-8, and 309793/2023-8, respectively. 
%\end{acknowledgments}

\appendix

\section{Canonical quantization, Weyl operators and Tomita-Takesaki modular theory}\label{App}

\subsection{The real massive scalar field in $(1+3)$ Minkowski spacetime}

We consider a free real scalar field of mass $m$ in $(1+3)$-dimensional Minkowski spacetime. Its quantized form, expressed via plane-wave decomposition, reads
	\begin{equation}\label{qf}
		\varphi(t,{\vec x}) = \int \! \frac{d^3 k}{(2 \pi)^3} \frac{1}{2 \omega_k} \left( e^{-ik_\mu x^\mu} a_k + e^{ik_\mu x^\mu} a^{\dagger}_k \right) \;, \qquad k_\mu x^\mu = \omega_k t - {\vec k} \cdot {\vec x} \;,
	\end{equation}
where $\omega_k = \sqrt{{\vec k}^2 + m^2}$ and the annihilation and creation operators satisfy the canonical commutation relations
	\begin{align}\label{eq:CCR}
		[a_k, a^{\dagger}_q] &= (2\pi)^3 \, 2\omega_k \, \delta^3(k - q), \\ \nonumber 
		[a_k, a_q] &= [a^{\dagger}_k, a^{\dagger}_q] = 0. 
	\end{align}

In a rigorous QFT framework, the scalar field is treated as an operator-valued distribution \cite{Haag:1992hx}, and thus must be smeared out with smooth, compactly supported test functions to yield well-defined operators in the Hilbert space. Given a smooth, compactly supported, test function $f(\vb*{x}) \in C_0^\infty(\mathbb{R}^{4})$, with $\vb*{x}=(t,{\vec x})$, the corresponding smeared field operator is defined as
\begin{align} 
		\varphi(f) = \int \! d^4\vb*{x} \; \varphi(\vb*{x}) f(\vb*{x}).
	\end{align}
The vacuum expectation value of two smeared field operators, given by  the two-point Wightman function, defines the Lorentz invariant inner product  $\langle f \vert g \rangle$:
\begin{align} \label{Inn}
		\langle f \vert g \rangle &= \langle 0 \vert \varphi(f) \varphi(g) \vert 0 \rangle =  \frac{i}{2} \Delta_{\rm PJ}(f,g) +  H(f,g),
	\end{align}
where $f, g \in C_0^\infty$, and $\Delta_{\text{PJ}}(f,g)$ and $H(f,g)$ denote the smeared Pauli–Jordan and Hadamard distributions, respectively. These are defined by
\begin{align}
		\Delta_{\rm PJ}(f,g) &=  \int \! d^4\vb*{x} d^4\vb*{y} f(\vb*{x}) \Delta_{\rm PJ}(\vb*{x}-\vb*{y}) g(\vb*{y}) \;,  \nonumber \\
		H(f,g) &=  \int \! d^4\vb*{x} d^4\vb*{y} f(\vb*{x}) H(\vb*{x}-\vb*{y}) g(\vb*{y})\;. \label{mint}
	\end{align}
with distributional kernels given explicitly by
\begin{eqnarray} 
\Delta_{\rm PJ}(\vb*{x}) & =& - \int \frac{d^3p}{(2\pi)^3} \frac{1}{\omega_k} \sin\left(p_\mu (x-x')^\mu \right)  \;, \nonumber \\
H(\vb*{x}) & = & \int \frac{d^3p}{(2\pi)^3} \frac{1}{2\omega_k} \cos\left( p_\mu (x-x')^\mu \right)  \;. \label{PJH}
\end{eqnarray}
Both  $\Delta_{\rm PJ}(\vb*{x})$  and $H(\vb*{x})$ are Lorentz-invariant. The Pauli-Jordan distribution  $\Delta_{\rm PJ}(\vb*{x})$  encodes the principle of causality, vanishing outside the light cone. Moreover, it is odd under $\vb*{x} \rightarrow -\vb*{x}$, whereas the Hadamard function $H(\vb*{x})$ is even. The commutator of the field operators is thus $\left[\varphi(f), \varphi(g)\right] = i \Delta_{\rm PJ}(f,g)$, ensuring that  $\left[\phi(f), \phi(g)\right] = 0,$ whenever the supports of $f$ and $g$ are spacelike separated. This compactly encodes the principle of micro-causality in relativistic field theory. \\\\The inner product $\langle f \vert g \rangle$ can be rewritten in momentum space as 
\begin{equation} 
\langle f \vert g \rangle = \int \frac{d^4p}{(2 \pi)^4} \; (2\pi) \theta(p^0)\delta(p^2 -m^2) f^{*}(p) g(p) \;, \label{fin}
\end{equation} 
 where $(f(p),g(p)$ stand for the Fourier transform of $(f(x),g(x)$: 
 \begin{equation} 
 f(p) = \int d^4x \; e^{i p_\mu x^\mu} f(x) \;. \label{ft}
 \end{equation}

\subsection{Basics of the Tomita-Takesaki modular theory} 

The Tomita-Takesaki modular theory is  a powerful framework for analyzing the Bell-CHSH inequality in Quantum Field Theory \cite{Summers:1987fn,Summers:1987squ,Summers:1987ze}. To set the stage, it is worth briefly reviewing some of the basic features of this elegant  theoretical structure. \\\\Let ${\cal O}$ stand for an open region of the Minkowski spacetime and let ${\cal M}({\cal O})$ be the space of test functions with support contained in $\cal O$: 
\begin{equation} 
	{\cal M}({\cal O}) = \{ f; \; {\rm such \; that} \; supp(f) \subseteq {\cal O} \}. \label{MO}
\end{equation}
One  introduces the symplectic complement of ${\cal M}({\cal O})$ as 
\begin{equation} 
	{\cal M'}({\cal O}) = \{ g; \; \Delta_{\rm PJ}(g,f) =0 \;\; \forall f \in {\cal M}({\cal O})\}. \label{MpO}
\end{equation}
In other words, ${\cal M'}({\cal O})$ is given by the set of all test functions for which the smeared Pauli-Jordan expression defined by Eq.~\eqref{mint} vanishes. The usefulness of the symplectic complement ${\cal M'}({\cal O})$ relies on the fact that it allows us to rephrase causality as 
\begin{equation} 
	\left[ \varphi(g) , \varphi(f) \right] = 0 \;, \;\; \forall \;\;g\in {\cal M'}({\cal O}) \;\;{\rm and} \;\;f \in {\cal M}({\cal O}). \label{MMM}
\end{equation}
The next step is that of introducing a von Neumann algebra  ${\cal F}({\cal M})$ of bounded operators supported in ${\cal O}$, equipped with a cyclic and separating state. As a concrete example of such an algebra, one may consider the von Neumann algebra generated by the Weyl operators, namely 
\begin{equation} 
{\cal F}({\cal M}) = {\rm span}\;\left\{  { W}_h \;, supp(h) \in {\cal M} \right\} \;, \label{vn}
\end{equation}  
where ${ W}_h$ stands for the unitary Weyl operator ${ W}_h = e^{i {\varphi}(h) }.$
%\begin{equation}
%{\cal A}_h = e^{i {\varphi}(h) }\;. \label{Weyl}
%\end{equation}
Using the relation $e^A \; e^B = \; e^{ A+B +\frac{1}{2}[A,B]}$,
%\begin{equation}
%e^A \; e^B = \; e^{ A+B +\frac{1}{2}[A,B]}, \label{exp_AB}
%\end{equation} 
valid for two operators $(A,B)$ commuting with $[A,B]$, one finds that the Weyl operators give rise to the following algebraic structure:
\begin{eqnarray}
{W}_h \;{ W}_h' & =  & e^{- \frac{1}{2} [{\varphi}(h), {\varphi}(h')] }\;{ W}_{(h+h')} = e^{ - \frac{i}{2} \Delta_{\textrm{PJ}}(h,h')}\;{ W}_{(h+h')},  \nonumber \\
{ W}^{\dagger}_h & = & { W}_{(-h)}, \label{algebra} 
\end{eqnarray} 
where $\Delta_{\textrm{PJ}}(h,h')$ is the smeared causal Pauli-Jordan expression \eqref{mint}. Setting $\varphi(h) = a_h + a^\dagger_h \;,$
%\begin{equation} 
%\varphi(h) = a_h + a^\dagger_h \;, \label{dec}
%\end{equation}
with $(a_h,a^\dagger_h)$ being the smeared annihilation and creation operators 
\begin{equation}
a_h = \int \! \frac{d^3 k}{(2 \pi)^3} \frac{1}{2 \omega_k} h^{*}(\omega_k,{\vec k}) a_k \;, \qquad a_h^\dagger = \int \! \frac{d^3 k}{(2 \pi)^3} \frac{1}{2 \omega_k} h(\omega_k,{\vec k}) a^\dagger _k \;, \qquad [a_h,a^{\dagger}_{h'}]= \langle h | h'\rangle \;, \label{smaad}
\end{equation}
it follows that the vacuum expectation value of the operator ${W }_h$ turns out to be
\begin{equation} 
\langle 0 \vert  {W}_h  \vert 0 \rangle = \; e^{-\frac{1}{2} {\lVert h\rVert}^2}, \label{vA}
\end{equation} 
where ${\lVert h\rVert}^2 \equiv \langle h | h \rangle$ and the vacuum state $\vert 0 \rangle$ is defined by $a_k \vert 0 \rangle=0, \forall k$. In particular, if $supp_f$ and $supp_g$ are spacelike separated, causality ensures that the Pauli-Jordan function vanishes. Thus, from the above properties, it follows the useful relation 
\begin{equation} 
\langle 0 \vert  {W}_f {W}_{g}  \vert 0 \rangle =  \langle 0 \vert {W}_{(f + g)} \vert 0 \rangle =
\; e^{-\frac{1}{2} {\lVert f+g \rVert}^2}. \label{vAhh}
\end{equation}
A very important property of the von Neumann algebra ${\cal F}({\cal M})$, Eq.~\eqref{vn}, generated by the Weyl operators is that, due to the Reeh-Schlieder theorem \cite{Haag:1992hx,Witten:2018zxz}, the vacuum state $|0\rangle$ is both cyclic and separating, meaning that: $i)$ the set of states $\{{W}_h \vert 0 \rangle$, $ {W}_h \in {\cal F}({\cal M})\}$ are {\it dense} in the Hilbert space; $ii)$ the condition $\{{W}_h \vert 0 \rangle = 0, {W}_h \in {\cal F}({\cal M})\}$, implies ${W}_h = 0$. \\\\In such a situation, one can apply the modular theory of Tomita-Takesaki \cite{Takesaki:1970aki,Bratteli:1979tw,Summers:2003tf,Guido:2008jk,Witten:2018zxz}, which will be presented for a generic von Neumann algebra ${\cal F}({\cal M})$ with a cyclic and separating state $|\omega\rangle$. To begin, it is helpful to remind the  notion  of {\it commutant} 
${\cal F'}({\cal M})$ of the  von Neumann algebra ${\cal F}({\cal M})$, namely   
\begin{equation} 
	{\cal F'}({\cal M}) = \{ w';  \;\;w'\;w = w \;w'\; \;\; \forall \; w \in {\cal F}({\cal M}) \}, \label{comm} 
\end{equation} 
{\it i.e.}, ${\cal F'}({\cal M})$ contains all elements which commute with each element of ${\cal F}({\cal M})$. Let us also state the  so-called Haag's duality \cite{Witten:2018zxz,Haag:1992hx,eckmann1973application}, ${\cal F'}({\cal M}) = {\cal F}({\cal M'}),$
%\begin{equation}
%	{\cal W'}({\cal M}) = {\cal W}({\cal M'}), \label{Hd}
%\end{equation}
namely, the commutant ${\cal F'}({\cal M})$ coincides with the elements of ${\cal F}({\cal M'})$ obtained by taking elements belonging to the symplectic complement ${\cal M'}$ of ${\cal M}$. This duality, to our knowledge, has only been proven in the case of free  fields \cite{eckmann1973application}. \\\\The 
Tomita-Takesaki construction makes use of  an anti-linear unbounded operator $S$ \cite{Bratteli:1979tw} acting on the von Neumann algebra ${\cal W}({\cal M})$ as 
\begin{align} 
	S \; w \vert \omega \rangle = w^{\dagger} \vert \omega \rangle, \qquad \forall w \in {\cal F}({\cal M})\;,  \label{TT1}
\end{align}  
from which it follows that $S \vert \omega \rangle = \vert \omega \rangle$ and $S^2 = 1$. Making use of the polar decomposition \cite{Bratteli:1979tw}
\begin{equation} 
S = J \; \Delta^{1/2} \;, \label{PD}
\end{equation} 
where $J$ is the anti-linear modular conjugation operator and $\Delta$ is the self-adjoint and positive modular operator, the following properties holds \cite{Takesaki:1970aki,Bratteli:1979tw,Summers:2003tf,Guido:2008jk,Witten:2018zxz} 
\begin{eqnarray} 
 J^{\dagger} & = & J \;, \qquad J^2 = 1 \;, \nonumber \\
 J \Delta^{1/2} \; J & = & \Delta^{-1/2} \;, \nonumber \\
 S^{\dagger} & = & J \Delta^{-1/2} \;, \qquad S^{\dagger} S^{\dagger} = 1 \;, \nonumber \\
\Delta & = & S^{\dagger} S \;, \qquad \Delta^{-1} = S S^{\dagger} \;. \label{TTP}
\end{eqnarray}
We can now state the renowned  Tomita-Takesaki theorem \cite{Takesaki:1970aki,Bratteli:1979tw,Summers:2003tf,Guido:2008jk,Witten:2018zxz}, namely:\\\ $i)$ $J \;{\cal F}({\cal M})\; J  =  {\cal F'}({\cal M})$ as well as $J \;{\cal F'}({\cal M}) \;J = {\cal F}({\cal M})$;\\\ $ii)$ there is  a one-parameter family of operators $\Delta^{it}$ with $t \in \mathbb{R}$ which leave ${\cal F}({\cal M})$ invariant, that is:
\begin{equation*}
    \Delta^{it} \; \;{\cal F}({\cal M}) \; \Delta^{-it} = \;{\cal F}({\cal M}).
\end{equation*}
This theorem has far reaching consequences and finds applications in many areas \cite{Summers:2003tf}. As far as the Bell-CHSH inequality is concerned, it provides a powerful way of obtaining Bob's operators from Alice's ones by means of the anti-unitary operator $J$. The construction goes as follows: One picks up two test functions $(f,f')\in {\cal M}({\cal O})$ and consider  the two Alice's operators $({ W}_f, { W}_{f'})$,
\begin{equation} 
 {W}_f = e^{i {\varphi}(f) } \;, \qquad  {W}_{f'} = e^{i {\varphi}(f') } \;. \label{AAop}
\end{equation} 
Thus, for Bob's operators, we write 
\begin{equation} 
J {W}_f J \;, \qquad J {W}_{f'} J \;. \label{Bobbop}
\end{equation} 
From the Tomita-Takesaki Theorem, it follows that the two set of operators $({W}_f, W_{f'})$ and $(J{W}_f J, J {W}_{f'}J)$ fulfill the necessary requirements, since $(J{W}_f J, J {W}_{f'}J)$ belong to the commutant ${\cal F'}({\cal M})$. Also, it is worth underlining that  the action of the operators $J$ and $\Delta$ may be lifted directly into the space of the test functions  \cite{eckmann1973application,rieffel1977bounded,Guido:2008jk}, giving rise to an analogue of the Tomita-Takesaki construction, namely,  
\begin{equation} 
J {W}_f J =  J e^{i {\varphi}(f) } J  \equiv e^{-i {\varphi}(jf) } \;, \label{jop}
\end{equation} 
where the test function $jf \in {\cal M'}$. Analogously, 
\begin{equation} 
\Delta^{1/2} {W}_f  \Delta^{-1/2} \equiv e^{i {\varphi}(\delta^{1/2}f) } \;. \label{dp}
\end{equation} 
The operators $(j,\delta)$ are such that \cite{eckmann1973application,rieffel1977bounded,Summers:1987squ,Guido:2008jk}
\begin{eqnarray}
s & = & j \delta^{1/2} \;, \qquad s^2=1 \;, \nonumber \\
j \delta^{1/2} j & = & \delta^{-1/2} \;, \nonumber  \\
s^{\dagger} & = & j \delta^{-1/2} \;, \qquad s^{\dagger} s^{\dagger} = 1 \;. \label{ssd}
\end{eqnarray}
The operator $j$ is anti-unitary, while $\delta$ is self-adjoint and positive. Moreover, from \cite{Summers:1987squ}, one learns that, in the case in which the region ${\cal O}$ is a wedge region of the Minkowski spacetime, the spectrum of the operator $\delta$ is the whole positive real line, {\it i.e.} $\log(\delta) =\mathbb{R}$. This result follows from the analysis of Bisognano-Wichmann \cite{Bisognano:1975ih} of the Tomita-Takesaki modular operator for wedge regions $({\cal O}_R,{\cal O}_L)$ 
\begin{equation}
{\cal O}_R =\left\{ (t,x,y,z)\;, x \ge |t| \right\} \;, \qquad  {\cal O}_L =\left\{ (t,x,y,z)\;, -x \ge |t| \right\} \;. \label{wedges}
\end{equation}
The wedge regions $({\cal O}_R,{\cal O}_L)$ are causal complement of each other and are left invariant by Lorentz boosts. In what follows, we shall always consider the region ${\cal O} $ to be a wedge region. For instance, one may figure out that Alice is located in the right  wedge ${\cal O}_R$, while Bob is in the left one ${\cal O}_L$. \\\\The operators $(s,s^{\dagger})$ have the physical meaning of projecting into the space ${\cal M}$ and its symplectic complement ${\cal M'}$, namely, one can show \cite{rieffel1977bounded, Summers:1987squ,Guido:2008jk}  that  a test function $f$ belongs to $\cal M$ if and only if
\begin{equation} 
s f = f \;. \label{sf}
\end{equation} 
Analogously, a test function $g$ belongs to the symplectic complement ${\cal M'}$ if and only if
\begin{equation} 
s^{\dagger} g = g \;. \label{sdg}
\end{equation} 
These properties enable us to construct a useful set of test functions for Alice and Bob. Following \cite{Summers:1987fn,Summers:1987squ}, Alice's test functions $(f,f')$ can be  specified as follows: Picking up the spectral subspace of $\delta$ specified by $[\lambda^2-\varepsilon, \lambda^2+\varepsilon ] \subset (0,1)$ and introducing  the normalized vector $\phi$ belonging to this subspace, one writes  
\begin{equation}
f  =\frac{1}{\sqrt{1-\lambda^2}} (1+s) \phi \;, \qquad f' = \frac{1}{\sqrt{1-\lambda^2}}(1+s) i \phi \;, 
\label{nmf}
\end{equation}
According to the setup outlined above, Eq.\eqref{nmf} ensures that $s f = f $ and $s f'= f' \;.$
%\begin{equation}
%s f = f  \;, \qquad s f'= f' \;.  \label{fafa}
%\end{equation}
Moreover, one checks  that $j\phi$ is orthogonal to $\phi$, {\it i.e.} $\langle \phi |  j\phi \rangle = 0$. In fact, from 
\begin{align} 
\delta^{-1} (j \phi) =  j (j \delta^{-1} j) \phi = j (\delta \phi), 
\label{orth}
\end{align}
it follows that the modular conjugation $j$ exchanges the spectral subspace $[\lambda^2-\varepsilon, \lambda^2+\varepsilon ]$ into $[1/\lambda^2-\varepsilon,1/ \lambda^2+\varepsilon ]$, ensuring that $\phi$ and $j \phi$ are orthogonal.   \\\\Concerning now the pair of Bob's test functions $(g,g')$, they are given by  
\begin{equation} 
g  =\frac{\lambda}{\sqrt{1-\lambda^2}} (1+s^\dagger) \phi \;, \qquad g' = -\frac{\lambda}{\sqrt{1-\lambda^2}}(1+s^\dagger) i \phi \;, 
\label{nmgg}
\end{equation}
meaning that, as required by the relativistic causality, $(g,g')$ belong to the symplectic  complement $\mathcal{M'}(\mathcal{O})$.  \\\\Finally,  it follows that \cite{Summers:1987fn,Summers:1987squ,Guimaraes:2024mmp}, in the limit $\varepsilon \rightarrow 0$,
\begin{align}
\vert\vert f \vert\vert^2  &= \vert\vert f'  \vert\vert^2 = \vert\vert g  \vert\vert^2 =\vert\vert g'  \vert\vert^2 = \frac{1+\lambda^2}{1-\lambda^2}
 \;, \nonumber \\
\langle f \vert g  \rangle &= - \langle f' \vert g'  \rangle=   \frac{2 \lambda}{1-\lambda^2} \;,  \nonumber \\
\langle f \vert g'  \rangle &=\langle f' \vert g  \rangle = 0 \;, \nonumber \\
\langle f \vert f'  \rangle &=\langle g \vert g'  \rangle = i \;. 
  \label{ssfl}
\end{align}

$\clubsuit$ $\clubsuit$ $\clubsuit$ $\clubsuit$

%%%%%%%%%%%%%%%%%%%%%%%%%%%%%%%%%%%%%%%%%%%%%%%%%%%%%%%%%%%%%%%%%%%%%%%%%%%%%%%%%%%%%%%%%%%%%%%%%%%%%%%%%%

\bibliography{refs}

\end{document}